\documentclass[useAMS,usenatbib]{mn2e}

\usepackage{graphicx}
\title[Lensing in Conformal Weyl Gravity]{Strong lensing as a test for Conformal Weyl Gravity}
\author[D. Cutajar and K. Zarb Adami]{Deandra Cutajar $^{1}$\thanks{
dcut0014@um.edu.mt} and Kristian Zarb Adami$^{1}$ $^{2}$\thanks{kristian.zarb-adami@um.edu.mt}\\
$^{1}$ Physics Department, University of Malta, Msida, MSD 2080, Malta\\
$^{2}$ Physics Department, University of Oxford, Oxford, OX1 3RH, United Kingdom}

\date{}

\begin{document}

\maketitle

\begin{abstract}
Conformal Weyl Gravity (CWG) predicts galactic rotation curves without invoking dark matter. This makes CWG an interesting candidate theory as an alternative to GR. This removal of the necessity of dark matter arises because the exact exterior solution in CWG for a static, spherically symmetric source provides a linear potential $\gamma r$, which can explain the observed galactic rotation curves with a value for $\gamma$ given by $\sim + 10^{-26} \mathrm{m}^{-1}$. Previous work has also shown that CWG can explain lensing observations, but with $\gamma\sim - 10^{-26} \mathrm{m}^{-1}$ in order to ensure converging rays of light rather than diverging ones. Even though different expressions for the angle of deflection have been derived in CWG in order to remove this inconsistency, this paper shows that the $\gamma$  required to fit the lensing data is several orders of magnitude higher than that required to fit the galactic rotation curves.
\end{abstract}

\begin{keywords}
 gravitational lensing, conformal weyl gravity, light bending, alternative theory of gravity, strong lensing, einstein radius
\end{keywords}

\section{Introduction}
Ever since the first indications of an invisible mass present in the Universe, the existence of dark matter remains an open question. Studies show that this matter could be composed of particles that do not emit light, but interact weakly with gravity. Despite the continuous search for particles that may make up dark matter, we still do not know whether such particles exist or another theory of gravity is needed to explain these phenomena. Gravitational lensing was predicted by Einstein's General Theory of Relativity (GR) and observed a few years later, nonetheless it provides a useful method to test alternative theories of gravity. \\
Without invoking dark matter but rather by modifying the expression for the gravitational potential, these theories aim to match the observed data with theoretical predictions. Thus such theories might give an explanation for phenomena, such as gravitational lensing without the need for dark matter.\\
Using a modified gravitational potential, Conformal Weyl gravity (CWG) shows remarkable predictions for the galactic rotation curves without the necessity of an extra invisible mass. However we cannot put aside dark matter and GR until we have more evidence that this theory proves the unnecessary presence of this additional mass. The objective of this paper is to test whether this theory also gives the same predictions for strong lensing. \\
Section 2 introduces the principle behind CWG and in Section 3 we discuss its predictions that fit observations for the galactic rotation curves. Section 4 focuses on gravitational lensing by providing a schematic explanation of the effect and explaining how the lens equation was obtained by Einstein himself, using simple trigonometry. Before proceeding to the paper's main contribution, we discuss previous studies on lensing and give an explanation for our use of different equations for the angle of deflection to test CWG in the strong lensing regime. We obtain an expression for Einstein's Radius (\textit{$R_E$}) in CWG by using the equation representing the deflection angle derived by Sultana and Kazanas (2010) and also that obtained by Cattani et.al. (2013). A data fitting algorithm was used to find $\gamma$ for which the numerical predictions match observations followed by a discussion on the results obtained in this study, before concluding in Section 5.. 
\section{Conformal Weyl Gravity}
CWG uses the principle of local conformal invariance of the space time manifold, in other words invariance under local conformal stretching \citep{Kazanas1989} \citep{Kazanas1991},
\begin{equation}
g_{\mu \nu} (x) \rightarrow \Omega^2 (x) g_{\mu \nu},
\end{equation}
where $g_{\mu \nu}$ is the space-time metric. The unique action of CWG is represented by 
\begin{equation}
I_w = - \alpha \int d^4x(-g)^{1/2} C_{\lambda \mu \nu \kappa}C^{\lambda \mu \nu \kappa},
\end{equation}
where $C_{\lambda \mu \nu \kappa}$ is the conformal Weyl tensor and $\alpha$ is a dimensionless coefficient. Eqn (2) leads to fourth order equations of motion for the gravitational field given as
\begin{equation}
- 2 \alpha W^{\mu\nu} = - 2 \alpha({C^{\lambda \mu \kappa \nu}}_{; \lambda \kappa} - \frac{1}{2}R_{\lambda \kappa} C^{\lambda \mu \kappa \nu}) = \frac{1}{2} T^{\mu \nu},
\end{equation}
for a source $T_{\mu \nu}$. Hence any vacuum solution for Einstein's field equation i.e. when $R_{\mu \nu}$ is zero, leads to a vacuum solution for Weyl gravity since $W_{\mu \nu}$ vanishes. In fact the exact exterior static and spherically symmetric vacuum solution for CWG is given by the metric \citep{Kazanas1989}
\begin{equation}
ds^2 = - B(r) dt^2 + \frac{dr^2}{B(r)}+r^2(d \theta ^2 + \sin^2\theta  d\phi ^2),
\end{equation}
where 
\begin{equation}
B(r) = 1- \frac{\beta(2-3 \beta \gamma)}{r}-3 \beta \gamma + \gamma r - k r^2,
\end{equation}
and $\beta$, $\gamma$ and \textit{k} are constants of integration. Outside the source, \textit{k} is related to the cosmological constant (\textit{$k \sim \frac{\Lambda}{3}$} where $\Lambda \sim$ $10^{-52}$ m$^{-2}$) in a Schwarzschild-de-Sitter background and $\beta$ = $\frac{GM}{c^2}$ (i.e. the geometric mass). Thus when $\gamma$ and \textit{k} tend to zero, Eqn (5) recovers the Schwarzschild solution in vacuum. \\
The key addition of CWG  is $\gamma$ since this describes the gravitational effect otherwise attributed to dark matter. As explained in Section 3, $\gamma$ succeeded in explaining this effect for galactic rotation curves where Solar system constraints require $ \vert \beta \gamma \vert \ll 1$.
\section{CWG and Galactic Rotation Curves}
While studying the galactic rotation curve, a Swiss Astronomer Zwicky (1937) proposed that more mass must be present than that observed.  Newtonian gravity failed to explain the observed rotational curves specifically for stars further away from the galactic centres. According to Newtonian mechanics, the rotational speeds should first increase with radius and then drop due to the absence of visible mass. Instead observations show that the curve reaches a limit which remains constant for higher radius. This could only be explained by invoking an additional invisible mass which was not detected \citep{Zwicky}. \\
Such observations were explained by CWG without dark matter by using the modified gravitational potential described by Eqn (5). As shown in Mannheim and Kazanas (1989) if \textit{$\gamma$r} is comparable to the Newtonian \textit{1/r} for a regular galactic scale (10 kpc), the equation would represent an increasing potential for \textit{r} $>$ 10 kpc;  constant for \textit{r} $\sim$ 10 kpc and Newtonian for \textit{r} $<$ 10 kpc. CWG fits rotation curves \citep{Mannheim1993}\citep{Mannheim1997} with the following relation
\begin{equation}
\gamma = \gamma_0 + \left(\frac{M}{M_{\sun}}\right) \gamma^{\star}
\end{equation}
where $\gamma_0 = 3.05 \times 10^{-30}$  cm$^{-1}$ and $\gamma^{\star} = 5.42 \times 10^{-41}$ cm$^{-1}$. This implies that for galactic rotation curves, dark matter is not essential and a modified gravitational potential may explain the observed phenomena. If Eqn (6) is applied to the clusters used in this analysis, where $M \sim 10^{13} M_{\sun}$, then $\gamma$ is found to be $10^{-26}$ m$^{-1}$, i.e. the inverse order of the Hubble length. Using a sample of galaxy clusters given by Tables \ref{sample} and \ref{results}, we test the lensing predictions for CWG. 
\section{CWG and Gravitational Lensing}
\begin{figure}
  \includegraphics[height=0.3 \textheight, width=0.4 \textwidth]{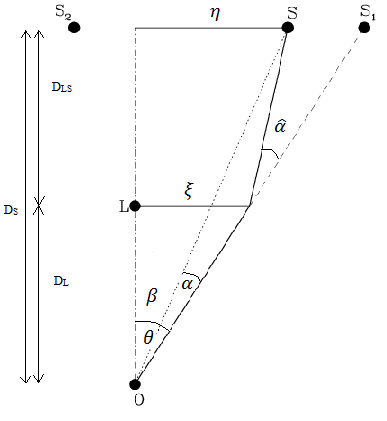}
 \caption{Gravitational Lensing by a Point Mass \citep{Wambsganss2001}.}\label{lens}
 \end{figure}

Strong gravitational lensing describes the bending of light in the presence of a gravitational field. Light is deflected as it passes through the gravitational potential of an intervening mass, a galaxy or a cluster of galaxies, between the source and the observer. The effect is similar to that caused by a lens. Fig \ref{lens} shows the lensing geometry where \textit{$D_S$}, \textit{$D_L$} and \textit{$D_{LS}$} are the angular diameter distances between the source and the observer, the lens and the observer and of the source as seen by the lens, respectively. $\theta_E$, the Einstein angle is related to the Einstein Radius shown in Eqn (7) \citep{Wambsganss2001}.
\begin{equation}
\xi_E = \theta_E  D_L.
\end{equation} 
$\hat{\alpha}$ is the angle of deflection and $\xi$ represents the impact parameter. Assuming the lens is spherically symmetric, then
\begin{equation}
\theta D_S = \beta D_S + \alpha(\theta) D_S.
\end{equation}
Thus, when placing the source on-axis behind the lens, i.e. $\beta$ = 0 in Eqn (8), the lens equation becomes
\begin{equation}
\theta D_S = \alpha (\theta) D_S,
\end{equation}
where
\begin{equation}
\alpha(\theta) = \frac{D_{LS}}{D_S} \hat{\alpha}.
\end{equation}
Substituting for $\alpha(\theta)$ we obtain
\begin{equation}
\theta_E = \frac{D_{LS}}{D_S} \hat{\alpha}. 
\end{equation}
Eqn (11) was derived using trigonometry and does not depend on a particular gravitational potential. Thus the same equation can be used for different theories of gravity where the $\hat{\alpha}$ is replaced accordingly and the distances should be computed with the respective expressions. Previous works on the angle of deflection in CWG shall be discussed in the next section, together with an explanation to why we apply the equations derived by Sultana and Kazanas (2010) and Cattani et.al. (2013) in this study.
\subsection{Previous Work}
Previous studies of lensing in CWG used the gravitational potential in Eqn (5) to find the deflection angle \citep{Edery1998}. The expression they obtained is,
\begin{equation}
\hat{\alpha}_{E\&P} = \frac{4 \beta}{\xi} - \gamma \xi.
\end{equation}
Pireaux (2004) derived a similar result, where the two middle terms obtained by Pireaux were neglected by Edery and Paranjape using the Solar system constraint $\vert \beta \gamma \vert \ll 1$. This leaves two main contributions to the angle of deflection; the first term which is the same as that in GR together with the last term $ - \gamma \xi$. As highlighted by both \citep{Edery1998} and \citep{Pireaux2004}, for the last term to bend light towards the source $\gamma < 0$.\\
A negative $\gamma$ in Eqn (12) would lead to a ``paradoxical result" as described in \citep{Sultana2010}. The linear relation between $\hat{\alpha}_{E\&P}$ and $\xi$ implies that ``the larger the light ray's impact parameter with respect to the lens, the larger the deflection angle" \citep{Sultana2010}. The deflection is an effect caused by an object's gravitational field. Since this effect decreases further away from the object, so does the deflection. In other words the closer to the object, the stronger the gravitational effect i.e. light should bend more as it travels close to the object than it would further away.\\
To address this problem, Sultana and Kazanas derived another expression for the bending of light from the same metric (4) using the gravitational potential in Eqn (5). In contrast to previous studies, Sultana and Kazanas do not obtained a `$+ \gamma \xi$' term in the expression for the angle of deflection. Instead the term is inversely proportional to the impact parameter. Such a result is expected and moreover contributes to the GR term which is also inversely proportional to the impact parameter. \\
Sultana and Kazanas (2010) followed the approach used by Rindler and Ishak (2008),  where the cosmological constant affects the bending of light. From the metric (4), the null geodesic equation is given as,
\begin{equation}
\frac{d^2 u}{d \phi} + u = 3 \beta \gamma u + \frac{3}{2}(2 - 3 \beta \gamma) \beta u^2 - \frac{\gamma}{2},
\end{equation}
where \textit{u} = $1/r$. One notes that the cosmological term vanishes in Eqn (13) but is introduced at a later stage. The final expression for the angle of deflection in CWG is \citep{Sultana2010}\footnote{S$\&$K is used to refer to the expression obtained by Sultana and Kazanas.},
\begin{equation}
\hat{\alpha}_{S\&K}=\frac{4 \beta}{\xi} - \frac{2 \beta^2 \gamma}{\xi} - \frac{k \xi^3}{2 \beta}.
\end{equation}
When replacing $\hat{\alpha}$ in Eqn (10) by (14) and using the following substitution $\xi = \theta D_L$, we obtain the quadratic equation,
\begin{equation}
\frac{k D_L^4}{2 \beta D} \theta^4 + \theta^2 - \frac{2 \beta }{D} (2 - \beta \gamma)  = 0.
\end{equation}
where $D \equiv \frac{D_L D_S}{D_{LS}}$. Solving for $\theta$ and taking the positive real solution, the Weyl angle ($\theta_{S\&K}$) in CWG is expressed by: 
\begin{equation}
\theta_{S\&K}^2 = -\frac{D \beta}{D_L^4 k} + \frac{\beta}{D_L^4 k}\sqrt{D^2 + 4 D_L^4 k (2 - \beta \gamma)}.
\end{equation}
Recent work by Cattani et.al. (2013) has shown that a negative $\gamma$ is not necessary for lensing. They derived an expression for the deflection angle comparable with that obtained by Sultana and Kazanas (Eqn 14) where in the former, the middle term has a positive contribution rather than negative. In comparison to the gravitational potential used by Sultana and Kazanas, Cattani et.al. use:
\begin{equation}
B(r) = \alpha - \frac{2 M}{r} + \gamma r - kr^2,
\end{equation}
where \textit{M} is the luminous mass and $\alpha = (1 - 6 M \gamma)^{1/2}$. Cattani et.al. follow the same approach as Sultana and Kazanas. They explain that if $M = \frac{\beta(2-3 \beta \gamma)}{2}$ and $\alpha = 1 - 3 \beta \gamma$ \citep{Cattani} were replaced in their gravitational potential (Eqn (17)), their equation for $\hat{\alpha}$ would be equivalent to Eqn (14). This approach led to a positive $\gamma$ term and a coefficient of 15 instead of -2 \footnote{Cat et.al. is used to refer to the expression obtained by Cattani et.al.},
\begin{equation}
\hat{\alpha}_{Cat\,et.al.}=\frac{4 M}{\xi} + \frac{15 M^2 \gamma}{\xi}.
\end{equation}
The Weyl Angle is thus expressed as \citep{Cattani} 
\begin{equation}
\theta_{Cat\,et.al.}^2 = \frac{4M + 15 M^2 \gamma}{D},
\end{equation}
where $D \equiv \frac{D_L D_S}{D_{LS}}$ as before.
\subsection{This Study}
\begin{figure*}
\centering
  \includegraphics[height=0.49 \textheight, width=1 \textwidth]{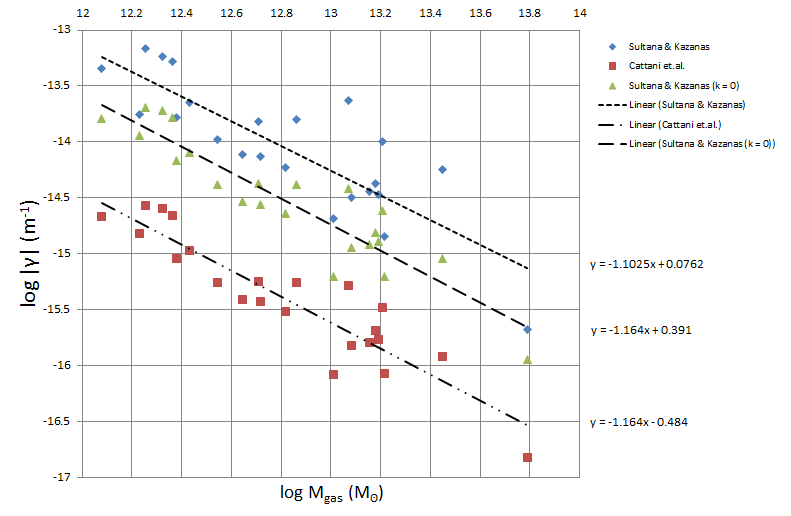}
  \caption{A plot of log $\gamma$ (m$^{-1}$) against log $M_{\rm gas}$ (M\sun).}\label{relgammas}
 \end{figure*}

In order to have an idea on the order of magnitude of $\gamma$, we represent $\hat{\alpha}_{CWG}$ as:
\begin{equation}
\hat{\alpha}_{CWG}=\frac{4 \beta}{\xi} + \frac{\varepsilon \beta^2 \gamma}{\xi} - \frac{k \xi^3}{2 \beta},
\end{equation}
where $\varepsilon = -2$ for the S$\&$K bend angle (Eqn (14)) and $\varepsilon = +15$ for the Cattani et.al. bend angle (Eqn (18)). Comparing the above equation with that derived for GR \citep{Rindler2008} in a Schwarzschild-de-Sitter background, the second term alters the effective Newtonian potential. As mentioned in the previous section, this term is inversely proportional to the impact parameter i.e. the light ray bends less as the ray travels at large distances from the object. The change in sign of $\gamma$ depends on the equation used for the analysis, such that Eqn (16) requires $\gamma < 0$ while $\gamma > 0$ is required for Eqn (19). \\
Using \textit{k} $\approx \frac{\Lambda}{3}$, the \textit{k}-term of Eqn (20) is the same as the $\Lambda$-term in \citep{Rindler2008}\citep{Ishak2008}, therefore we can equate the first two terms of Eqn (20) with the expression for $\hat{\alpha}$ in GR (when $\Lambda$ = 0), i.e.
\begin{equation}
\frac{4 \beta}{\xi} + \frac{\varepsilon \beta^2 \gamma}{\xi} = \frac{4 M}{R}.
\end{equation} 
Thus the second term of Eqn (21) is added to the first where \textit{M} (on the RHS) is the total geometric mass. In order to show how $\gamma$ should behave for lensing, the RHS of Eqn (21) is re-written as follows:
\begin{equation}
\frac{4M}{R} \rightarrow \frac{4M_{baryonic \, matter}}{R} + \frac{4M_{dark \, matter}}{R}. 
\end{equation}
In CWG, $\beta$ is only related to the baryonic matter so that the second terms should be equivalent leading to
\begin{equation}
\frac{\varepsilon \beta^2 \gamma}{\xi} = \frac{4M_{dark \, matter}}{R}.
\end{equation}
Now $\xi \simeq$ \textit{R}, which leaves the most important relation for this analysis, 
\begin{equation}
\beta^2 \gamma \approx \varepsilon' M_{dark \, matter}.
\end{equation}
From the above relation one notes that the Solar system constraint $\vert \beta \gamma \vert \ll 1$ does not hold for gravitational lensing, unless $\varepsilon'$ is of the order of $\approx 10^{-12}$. In fact Eqn (24) predicts that $\gamma$ should have the negative order of $\beta$ or a few orders less, so that the LHS has the same order of magnitude as the RHS i.e. $\gamma$ is inversely proportional to $\beta$ (the baryonic mass). This observation is clear since the dark matter mass in galaxies and clusters of galaxies, is either of the same order of magnitude as that of the baryonic matter mass or a few orders higher. In other words:
\begin{equation}
\gamma \approx \varepsilon' \frac{M_{dark \, matter}}{\beta^2}.
\end{equation}
As pointed out earlier in this section, the change in sign of $\gamma$ depends on the equation used and we proceed to confront this analysis with observations using the sample of galaxy clusters mentioned earlier. Eqns (16) and (19) were used to fit for a $\gamma$ which best explains lensing observations. The angular diameter distances were obtained using $d_A = \frac{d_L}{(1 + z)^2}$ \citep{Varieschi2011}, where the respective expression for the luminosity distance ($d_L$) in CWG \citep{Diaferio2011} \citep{Mannheim2003} \citep{Speirits2006} is given as,
\begin{equation}
d_L = \frac{c(1+z)^2}{q_0 H_0}\left[\left(1 + q_0 - \frac{q_0}{(1+z)^2}\right)^{\frac{1}{2}} - 1 \right].
\end{equation}
$q_0$ represents the deceleration parameter for which value was found to be $\sim -0.37$ \citep{Mannheim2003a} \citep{Diaferio2011} and $H_0$ is the Hubble constant. The redshift of the background sources (\textit{$z_{arc}$}) were also included in Tables \ref{sample} and \ref{results}. The equations derived by Edery and Paranjape (1998), Sultana and Kazanas (2010) and Cattani et.al. (2013) were used for clusters A 370 \citep{Richard}, A1689 \citep{Morandi}, A2163 and A2218 \citep{Makino} shown in Table \ref{sample}, so as to highlight the different behaviour of $\gamma$ arising from different expressions of the deflection angle.\\
In the analysis for $\theta_{S\&K}$, \textit{k} was taken to be constant and of the same order of the cosmological constant $\Lambda \sim$ $10^{-52}$ m$^{-2}$. In \citep{Ishak2008}\citep{Rindler2008}, the authors explain that the cosmological contribution of $\Lambda$ is small, nonetheless the \textit{k}-term of Eqn (14) was not discarded but kept for the expression of the angle in $\theta_{S\&K}$ (Eqn (16)).\\
Our fitting procedure for each arc computes $\beta$ ($ = \frac{GM}{c^2}$) for each respective enclosed gas mass ($M_{\rm gas}$), and then adjusts $\gamma$ until the Weyl angle (Eqns (16) \& (19)) matches the observed arc size. The resulting $\gamma$ values for a sample of clusters taken from \citep{Wu2000} (with $H_0$ = 50 kms$^{-1}$Mpc$^{-1}$) are given for each arc in Table \ref{results}. Fig 2 shows the resulting log $\vert \gamma \vert$ values plotted against the enclosed gas mass, log $M_{\rm gas}$, as obtained from Eqns (16) \& (19). Two series were plotted for Eqn (16) \citep{Sultana2010}, to show that the \textit{k}-term has a little effect on log $\vert \gamma \vert$.\\
In fact Fig \ref{relgammas} shows how small the contribution of the k-term is to the bending of light which was pointed out by  \citep{Rindler2008} \citep{Ishak2008}, where they discuss the small contribution of $\Lambda$ (\textit{$k \sim \frac{\Lambda}{3}$}). If this was not the case, then the ratio $\frac{\gamma_{S\&K}}{\gamma_{Cat\,et.al.}}$ would be greater than 10. \\
The expected inverse relation between $\gamma$ and $M_{gas}$ is also evident in Fig \ref{relgammas}. For each series we add a trend line to the points and their respective equations are shown next to the end of the line. In fact all three equations show a negative gradient, which supports our expectations.  The small contribution of the \textit{k}-term discussed above is shown again from the first data series, where $\gamma$ and $M_{gas}$ are still inversely related, however with a different slope from the other two lines. The y-intercept of the S$\&$K equations shows that including the \textit{k}-term would result in a higher value of $\gamma$ than that obtained in the other series. Therefore higher fits for \textit{k} \citep{Mannheim2012} would result in a larger negative term in Eqn (20) and therefore a larger positive $\gamma$ term than that obtained in this paper will necessary to account for its contribution. \\
 \begin{table*}
\centering
 \begin{minipage}{140mm}
  \caption{A sample of clusters.}\label{sample}
  \begin{tabular}{@{}lllrrrrrrrr@{}}
  \hline
Cluster & \textit{$z_L$} & $M_{gas}$ & $H_0$ & $\theta_{obs}$ & $\beta_{gas}$ \footnote{Geometric mass}  & $M_{lens}$ & \textit{$z_{arc}$} &  $\gamma_{E\&P}$   & $\gamma_{S\&K}$    & $\gamma_{Cat\,et.al.}$ \\
        &   & $10^{13}$ M\sun & kms$^{-1}$Mpc$^{-1}$  & \arcsec &  $10^{15}$ m    & $10^{13}$ M\sun   &   &$10^{-26}$ m$^{-1}$ & $10^{-14}$ m$^{-1}$& $10^{-15}$ m$^{-1}$    \\

 \hline
 A 370   & 0.375 & $0.73_{-0.06}^{+0.07}$ & 70 & 39.0 & 10.78 & 29.0 & 0.725 & -5.65 & - 3.81 & 1.23\\
 A1689   & 0.183 & $1.56_{-0.03}^{+0.03}$ & 70 & 45.0 & 23.04 & 36.0 & 1.000 & -5.50 & - 0.16 & 0.12    \\
 A2163   & 0.201 & $0.24_{-0.02}^{+0.02}$ & 50 & 15.6 &  3.54 &  4.1 & 0.728 & -4.18 & - 2.15 & 1.09\\
 A2218   & 0.175 & $0.18_{-0.01}^{+0.01}$ & 50 & 20.8 &  2.66 &  6.2 & 0.702 & -4.69 & - 8.79 & 3.09\\

\hline
\end{tabular}
\end{minipage}
\end{table*}

\begin{table*}
\centering
 \begin{minipage}{140mm}
  \caption{A sample of clusters \citep{Wu2000} showing constraints on $\gamma$.}\label{results}
  \begin{tabular}{@{}lllrrrrrrr@{}}
  \hline
Cluster & \textit{$z_L$} & $M_{gas}$ & $R_{obs}$ & $\beta_{gas}$ \footnote{Geometric mass}  & $M_{lens}$ & \textit{$z_{arc}$} & $\gamma_{S\&K}$    & $\gamma_{Cat\,et.al.}$ \\
        &   & $10^{13}$ M\sun  & kpc &  $10^{15}$ m    & $10^{13}$ M\sun   &   & $10^{-14}$ m$^{-1}$& $10^{-15}$ m$^{-1}$    \\

 \hline
 A 370   & 0.373 & $2.81_{-0.23}^{+0.25}$ & 350.0 & 41.50  & 130.0 & 1.3   & -0.57 & 0.12 \\    
 A 963   & 0.206 & $0.27_{-0.02}^{+0.02}$ &  80.0 &  1.77  &   6.0 & 0.711 & -2.30 & 1.08\\
         &       & $0.12_{-0.01}^{+0.01}$ &  51.7 &  3.99  &   2.5 & -     & -4.50 & 2.16\\
 A1689   & 0.181 & $1.56_{-0.03}^{+0.03}$ & 183.0 & 23.04  &  36.0 & -     & - 0.34 & 0.17\\
 A2218   & 0.171 & $0.21_{-0.01}^{+0.01}$ &  84.8 &  3.10  &   5.7 & 0.515 & -5.80 & 2.55\\
         &       & $1.61_{-0.04}^{+0.04}$ & 260.0 & 23.78  &  27.0 & 1.034 & -1.00 & 0.33\\ 
 A2219   & 0.228 & $0.35_{-0.02}^{+0.02}$ &  79.0 &  5.17  &   5.6 & -     & -1.10 & 0.56\\
         &       & $0.66_{-0.04}^{+0.04}$ & 110.0 &  9.75  &  16.0 & -     & -0.59 & 0.30\\
 A2390   & 0.228 & $1.43_{-0.21}^{+0.20}$ & 177.0 & 21.12  &  25.4 & 0.913 & -0.36 & 0.16\\
 A2744   & 0.308 & $0.51_{-0.07}^{+0.06}$ & 119.6 &  7.53  &  11.4 & -     & -1.50 & 0.57\\
 C10500  & 0.327 & $1.21_{-0.15}^{+0.14}$ & 150.0 & 17.87  &  19.0 & -     & -0.31 & 0.15\\
 MS0440  & 0.197 & $0.23_{-0.03}^{+0.03}$ &  89.0 &  3.40  &   8.9 & 0.530 & -5.30 & 2.20\\
 MS0451  & 0.539 & $1.51_{-0.05}^{+0.05}$ & 190.0 & 22.30  &  52.0 & -     & -0.42 & 0.21\\
 MS1008  & 0.306 & $1.18_{-0.16}^{+0.15}$ & 260.0 & 17.43  &  61.0 & -     & -2.30 & 0.52\\
 MS1358  & 0.329 & $1.02_{-0.12}^{+0.11}$ & 121.0 & 15.06  &   8.3 & 4.92  & -0.20 & 0.08\\
 MS1455  & 0.257 & $0.52_{-0.02}^{+0.02}$ &  98.0 &  7.68  &   8.6 & -     & -0.74 & 0.37\\
 MS2137  & 0.313 & $0.44_{-0.04}^{+0.05}$ &  87.4 &  6.50  &   7.1 & -     & -0.77 & 0.39\\
 PKS0745 & 0.103 & $0.17_{-0.01}^{+0.01}$ &  45.9 &  2.51  &   3.0 & 0.433 & -1.80 & 1.51\\
 RXJ1347 & 0.451 & $6.15_{-0.48}^{+0.47}$ & 240.0 & 90.82  &  42.0 & -     & -0.02 & 0.02\\  
            
\hline
\end{tabular}
\end{minipage}
\end{table*}

\section{Conclusion}

Previous studies showed that $\gamma$ of different signs is required for gravitational lensing in CWG \citep{Edery1998} \citep{Pireaux2004}, but is of the same order of magnitude as found in the study for the rotational velocities of galaxies. However, the deflection angle that led to such conclusions increased as $\gamma \xi$ rather than decreased with impact parameter $\xi$. This problem was addressed by Sultana and Kazanas (2010) who derived another expression for the deflection angle in CWG and recently also by Cattani et.al. (2013). In this study we used the latter two equations to understand how $\gamma$ behaves when the angle of deflection is inversely proportional to the impact parameter. \\
For CWG to fit galaxy rotation curves, we find that $\gamma$ is positive. However in the expression for the angle of deflection (Eqn (14)), as noted by previous studies \citep{Edery1998} \citep{Pireaux2004}, $\gamma < 0$ is required to fit lensing observations. Despite their agreement on the sign, this work disagrees on the order of magnitude for $\gamma$. Using Eqn (14), $\gamma$ turns out to be approximately $10^{12}$ orders higher than that obtained by Edery and Paranjape (1998). A similar behaviour of $\gamma$ was obtained using Cattani et.al.'s (2013) expression for the angle of deflection with $\gamma > 0$. \\
The gravitational potential (Eqn (5)) used to derive the equations for the angle of deflection represents the exact exterior solution for a static, spherically symmetric source. One could argue that the analysis should include the interior and the exterior solution in CWG \citep{intext}. Hence, light travelling in the vicinity of an intermediate mass between the source and an observer, necessarily passes through an exterior mass distribution and not through vacuum. Such scenarios would have two types of deflection to be considered: the bending of light caused by the interior mass acting as lens bending light towards it, and the divergence of the light ray away from the lens caused by the exterior distribution. Thus $\gamma$ should account for the effect otherwise attributed to dark matter, and also for the divergence caused by the exterior mass distribution. If the effect of the exterior distribution is small, then one would obtain similar results for $\gamma$ as obtained in this work. This further confirms the result of this paper to show that for lensing, $\gamma$ has to be several orders of magnitude higher than that obtained from galactic rotation curve.

\section*{Acknowledgements}

We would like to thank J. Sultana and J.L. Said for the helpful discussions throughout this work. We also thank the Reviewer for reading the manuscript and commenting on valuable aspects of our research which improved the interpretation of our results.
\bibliographystyle{mn2e}
\bibliography{Thesis}
\end{document}